# Sex and Coevolution

Larry Bull

Computer Science Research Centre

Department of Computer Science & Creative Technologies

University of the West of England, Bristol UK

Larry.Bull@uwe.ac.uk

**Abstract**

It has been suggested that the fundamental haploid-diploid cycle of eukaryotic sex exploits a rudimentary form of the Baldwin effect. This paper uses the well-known NKCS model to explore the effects of coevolution upon the behaviour of eukaryotes. It is shown how varying fitness landscape size, ruggedness and connectedness can vary the conditions under which eukaryotic sex proves beneficial over asexual reproduction in haploids in a coevolutionary context. Moreover, eukaryotic sex is shown to be more sensitive to the relative rate of evolution exhibited by its partnering species than asexual haploids.

**Introduction**

Coevolution refers to the effects upon the evolutionary behaviour of one species by the one or more other species with which it interacts. At an abstract level coevolution can be considered as the coupling together of the fitness landscapes of the interacting species. Hence the adaptive moves made one species in its fitness landscape causes deformations in the fitness landscapes of its coupled partners. In this paper Kauffman and Johnsen's [1992] NKCS model, which allows for the systematic alteration of various aspects of a coevolving environment, particularly the degree of landscape ruggedness and connectedness, is used to explore the coevolutionary behaviour of sexual diploid species. Eukaryotic sex is here defined as successive rounds of syngamy and meiosis in a haploid-diploid lifecycle. It has been suggested that the emergence of a haploid-diploid cycle enabled the exploitation of a rudimentary form of the Baldwin effect [Baldwin, 1896] and that this provides an underpinning explanation for all the observed forms of sex [Bull, 2017]. The Baldwin effect is here defined as the existence of phenotypic plasticity that enables an organism to exhibit a significantly different (better) fitness than its genome directly represents. Over time, as evolution is guided towards such regions under selection, higher fitness alleles/genomes which rely less upon the phenotypic plasticity can be discovered and become assimilated into the population.

Key to the new explanation for the evolution of sex in eukaryotes is to view the process from the perspective of the constituent haploids. A diploid organism may been seen to simultaneously represent two points in the underlying haploid fitness landscape. The fitness associated with those two haploids is therefore the fitness achieved in their combined form as a diploid; each haploid genome will have the same fitness value and that will almost certainly differ from that of their corresponding haploid organism due to the interactions between the two genomes. That is, the effects of haploid genome combination into a diploid can be seen as a simple form of phenotypic plasticity for the individual haploids before they revert to a solitary state during reproduction. In this way evolution can be seen to be both assigning a single fitness value to the *region* of the landscape between the two points represented by a diploid's constituent haploid genomes and altering the shape of the haploid fitness landscape. In particular, the latter enables the landscape to be smoothed under a rudimentary Baldwin effect process [Hinton & Nowlan, 1987].

Numerous explanations exist for the benefits of recombination (eg, [Bernstein and Bernstein, 2010]) but the role becomes clear under the new view: recombination facilitates genetic assimilation within the simple form of the Baldwin effect. If the haploid pairing is beneficial and the diploid is chosen under selection to reproduce, the recombination process can bring an assortment of those partnered genes together into new haploid genomes. In this way the fitter allele values from the pair of partnered haploids may come to exist within individual haploids more quickly than the under mutation alone (see [Bull, 2017] for full details).

All previous known studies of coevolution with the NKCS model have used asexual haploid species. This paper explores the behaviour of sexual diploid species in the model both to compare with the typical behaviour seen with asexual haploids and to determine where sex is beneficial within a coevolutionary context. Results show sexual diploids species generally appear more sensitive to coevolution than asexual haploids.

**The NK and NKCS Models**

Kauffman and Levin [1987] introduced the NK model to allow the systematic study of various aspects of fitness landscapes (see [Kauffman, 1993] for an overview). In the standard model, the features of the fitness landscapes are specified by two parameters: $N$, the length of the genome; and $K$, the number of genes that has an effect on the fitness contribution of each (binary) gene. Thus increasing $K$ with respect to $N$ increases the epistatic linkage, increasing the ruggedness of the fitness landscape. The increase in epistasis increases the number of optima, increases the steepness of their sides, and decreases their correlation. The model assumes all intragenome interactions are so complex that it is only appropriate to assign random values to their effects on fitness. Therefore for each of the possible $K$ interactions a table of $2^{(K+1)}$ fitnesses is created for each gene with all entries in the range 0.0 to 1.0, such that there is one fitness for each combination of traits (Figure 1). The fitness contribution of each gene is found from its table. These fitnesses are then summed and normalized by $N$ to give the selective fitness of the total genome.

Kauffman [1993] used a mutation-based hill-climbing algorithm, where the single point in the fitness space is said to represent a converged species, to examine the properties and evolutionary dynamics of the NK model. That is, the population is of size one and a species evolves by making a random change to one randomly chosen gene per generation. The "population" is said to move to the genetic configuration of the mutated individual if its fitness is greater than the fitness of the current individual; the rate of supply of mutants is seen as slow compared to the actions of selection. Ties are broken at random. Figure 2 shows example results. All results reported in this paper are the average of 10 runs (random start points) on each of 10 NK functions, ie, 100 runs, for 20,000 generations. Here $0 \leq K \leq 15$, for $N=20$ and $N=100$.

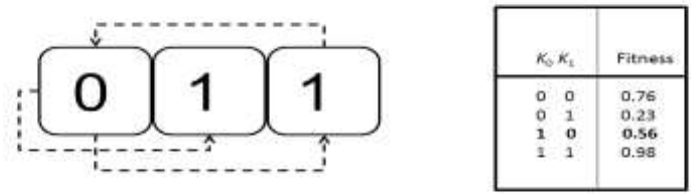

Figure 1: An example NK model ($N=3$, $K=1$) showing how the fitness contribution of each gene depends on $K$ random genes (left). Therefore there are $2^{(K+1)}$ possible allele combinations per gene, each of which is assigned a random fitness. Each gene of the genome has such a table created for it (right, centre gene shown). Total fitness is the normalized sum of these values.

Figure 2 shows examples of the general properties of adaptation on such rugged fitness landscapes identified by Kauffman (eg, [1993]), including a "complexity catastrophe" as $K \rightarrow N$. When $K=0$ all genes make an independent contribution to the overall fitness and, since fitness values are drawn at random between 0.0 and 1.0, order statistics show the average value of the fit allele should be 0.66. Hence a single, global optimum exists in the landscape of fitness 0.66, regardless of the value of $N$. At low levels of $K$ ($0<K<8$), the landscape buckles up and becomes more rugged, with an increasing number of peaks at higher fitness levels, regardless of $N$. Thereafter the increasing complexity of constraints between genes means the height of peaks typically found begin to fall as $K$ increases relative to $N$: for large $N$, the central limit theorem suggests reachable optima will have a mean fitness of 0.5 as $K \rightarrow N$. Figure 2 shows how the optima found when $K>6$ are significantly lower for $N=20$ compared to those for $N=100$ (T-test, $p<0.05$).

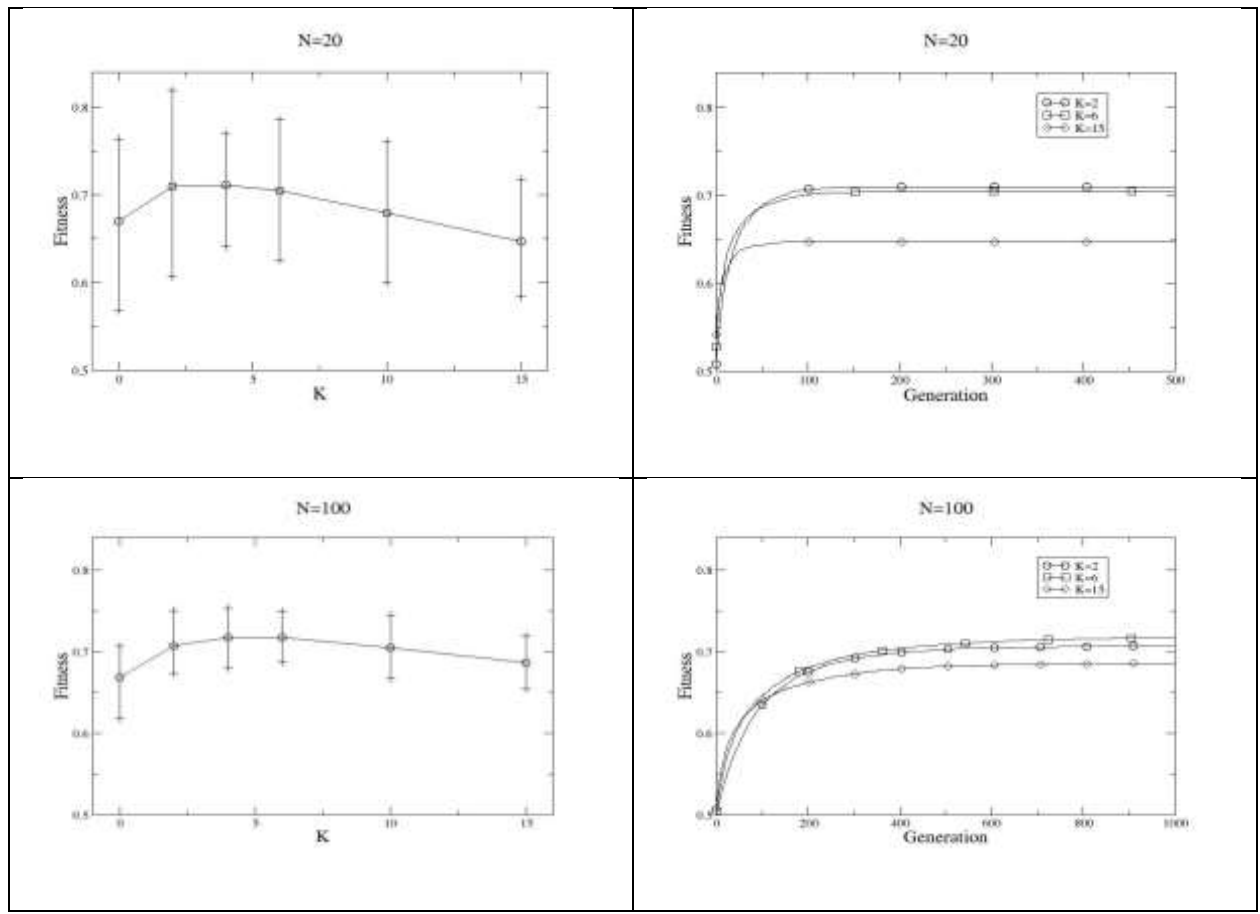

Figure 2: Showing typical behaviour and the fitness reached after 20,000 generations on landscapes of varying ruggedness ($K$) and length ($N$). Error bars show min and max values.

Kauffman and Johnsen [1992] subsequently introduced the abstract NKCS model to enable the study of various aspects of *co*evolution. Each gene is also said to depend upon $C$ randomly chosen traits in each of the other $S$ species with which it interacts. The adaptive moves by one species may deform the fitness landscape(s) of its partner(s). Altering $C$, with respect to $N$, changes how dramatically adaptive moves by each species deform the landscape(s) of its partner(s). Again, for each of the possible $K+(SxC)$ interactions, a table of $2^{(K+(SxC)+1)}$ fitnesses is created for each gene, with all entries in the range 0.0 to 1.0, such that there is one fitness for each combination of traits. Such tables are created for each species (Figure 3).

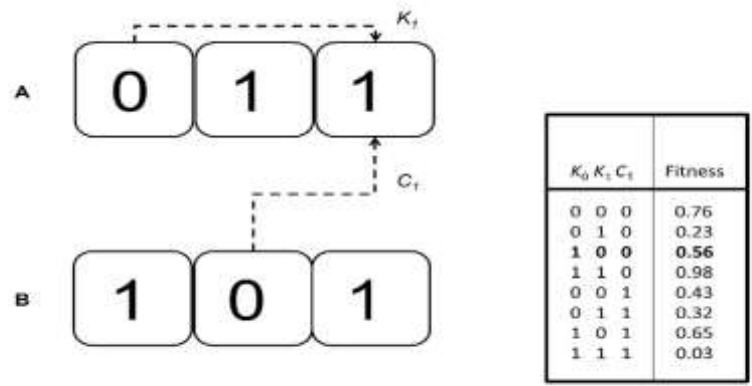

Figure 3: The NKCS model: Each gene is connected to $K$ randomly chosen local genes and to $C$ randomly chosen genes in each of the $S$ other species. A random fitness is assigned to each possible set of combinations of genes. These are normalised by $N$ to give the fitness of the genome. Connections and table shown for one gene in one species for clarity.

Figure 4 shows example results for one of two coevolving species where the parameters of each are the same and hence behaviour is symmetrical. All results reported in this paper are the average of 10 runs (random start points) on each of 10 NKCS functions, ie, 100 runs, for 20,000 generations. Here $0 \leq K \leq 10$, $1 \leq C \leq 5$, for $N=20$ and $N=100$. When $C=1$, Figure 4 shows examples of the general properties of adaptation on such fitness landscapes identified above in the NK model (where $C=0$). Figure 4 further shows how increasing the degree of connectedness ($C$) between the two landscapes causes fitness levels to fall significantly (T-test, $p<0.05$) when $C \geq K$ for $N=20$. That is, as $K \rightarrow N$ a high number of peaks of similar height typically exist in each of the fitness landscapes and so the effects of switching between them under the influence of $C$ is reduced since each landscape is very similar. Note this change in behaviour around $C=K$ was suggested as significant in [Kauffman, 1993], where $N=24$ was used throughout. However, Figure 4 also shows how with $N=100$ fitness *always* falls significantly with increasing *C* (T-test, *p*<0.05), regardless of *K*. That is, it might be concluded that more complex organisms (>*N*) appear more sensitive to landscape coupling (>*C*).

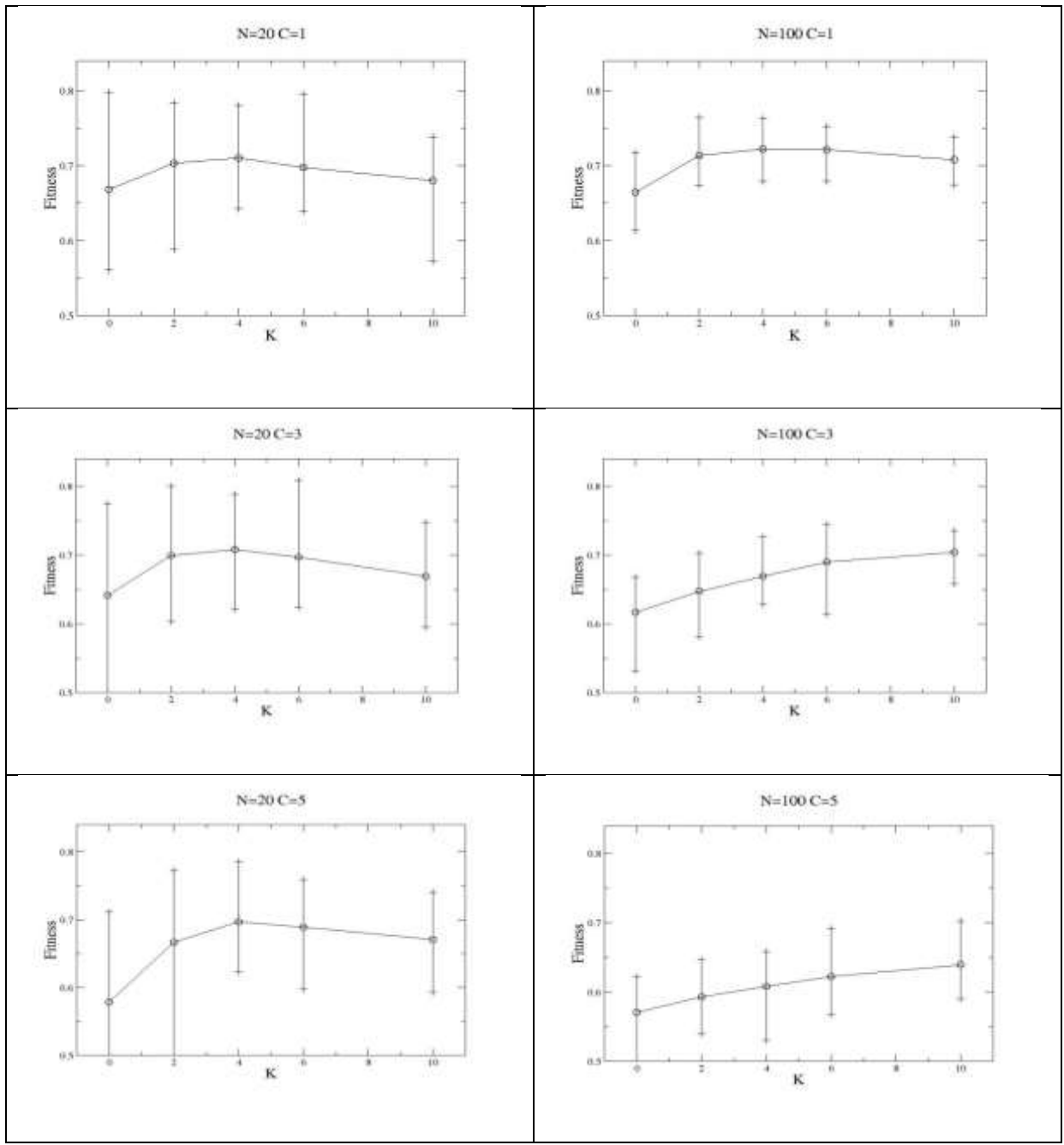

Figure 4: Showing the fitness reached after 20,000 generations on landscapes of varying ruggedness (*K*), coupling (*C*), and length (*N*). Error bars show min and max values.

# Sex in the NK and NKCS Models

As discussed in [Maynard-Smith & Szathmary, 1995, p150], the first step in the evolution of eukaryotic sex was the emergence of a haploid-diploid cycle, probably via endomitosis, then simple syngamy or one-step meiosis, before two-step meiosis with recombination. Following [Bull, 2017], the NK model can be extended to consider aspects of the evolution of sexual diploids. Firstly, each individual contains two haploid genomes of the form described above for the standard model. The fitness of an individual is here simply assigned as the average of the fitness of each of its constituent haploids. These are initially created at random, as before. Two-step meiosis with recombination is here implemented as follows: on each generation the diploid individual representing the converged population is copied twice to create two offspring. In each offspring, each haploid genome is copied once, a single recombination point is chosen at random, and non-sister haploids are recombined. One of the four resulting haploids in each offspring individual is chosen at random. Finally, a random gene in each chosen haploid is mutated. The resulting pair of haploids forms the new diploid offspring to be evaluated (Figure 5).

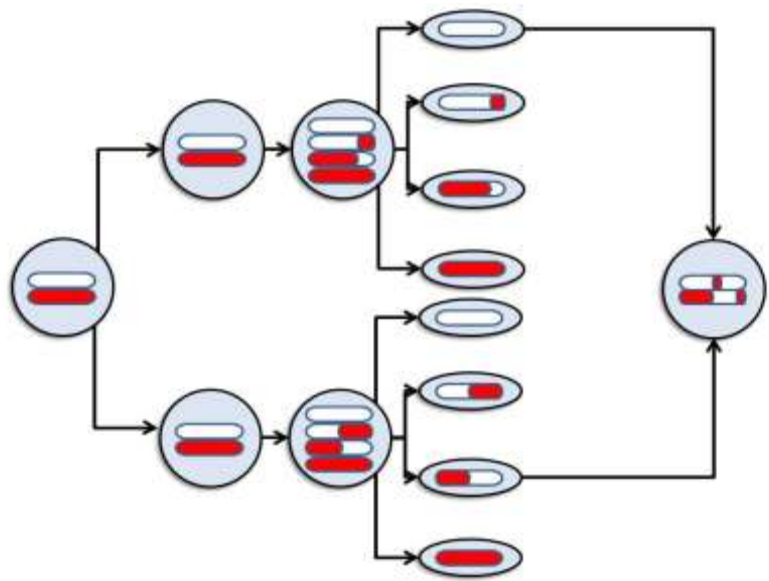

Figure 5: Showing the haploid-diploid cycle with meiosis as implemented with a converged population here.

Figure 6 shows examples of the general properties of adaptation in the NK model of rugged fitness landscapes for diploid organisms evolving via two-step meiosis with recombination. When $N=20$ the fitness level reached is significantly lower than for $N=100$ for $K>4$ (T-test, $p<0.05$), as is seen in the traditional haploid case due to the effects of the increased landscape complexity. Following [Bull, 2017], it can be seen that fitness levels are always higher than the equivalent haploid case (Figure 2) when $K>0$ due to the Baldwin effect, as discussed above (T-test, $p<0.05$).

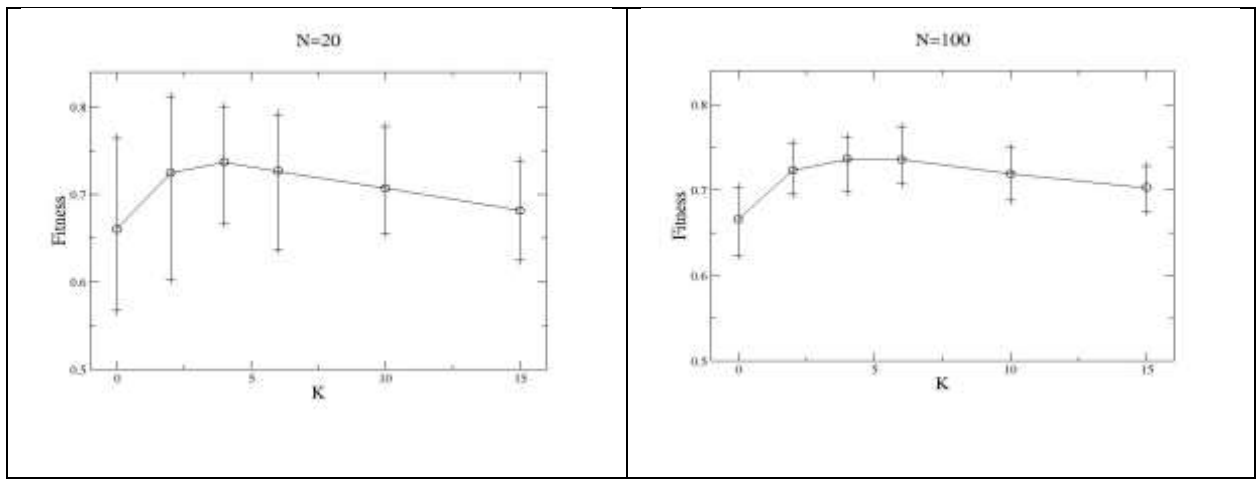

Figure 6: Showing typical behaviour and the fitness reached after 20,000 generations on landscapes of varying ruggedness ($K$) and length ($N$) for sexual diploids.

Figure 7 shows examples of extending the NKCS model such that one of a pair of species is a sexual diploid with other an asexual haploid. Comparing to Figure 4, with $C=1$ and $N=20$ or $N=100$, an increase in fitness compared to the equivalent asexual haploid can be seen for $K>0$ (T-test, $p<0.05$), presumably due to the Baldwin effect again being exploited as in the NK model (where $C=0$). When $C=3$ and $N=20$, sexual diploidy results in an increase in fitness over the haploid when $K>4$ (T-test, $p<0.05$) but it is always worse with $N=100$ (T-test, $p<0.05$). Conversely, with $C=5$ and $N=20$, sexual diploidy results in a lower fitness compared to the haploid when $K<6$ (T-test, $p<0.05$), with no significant difference for all $K$ when $N=100$ (T-test, $p≥0.05$).

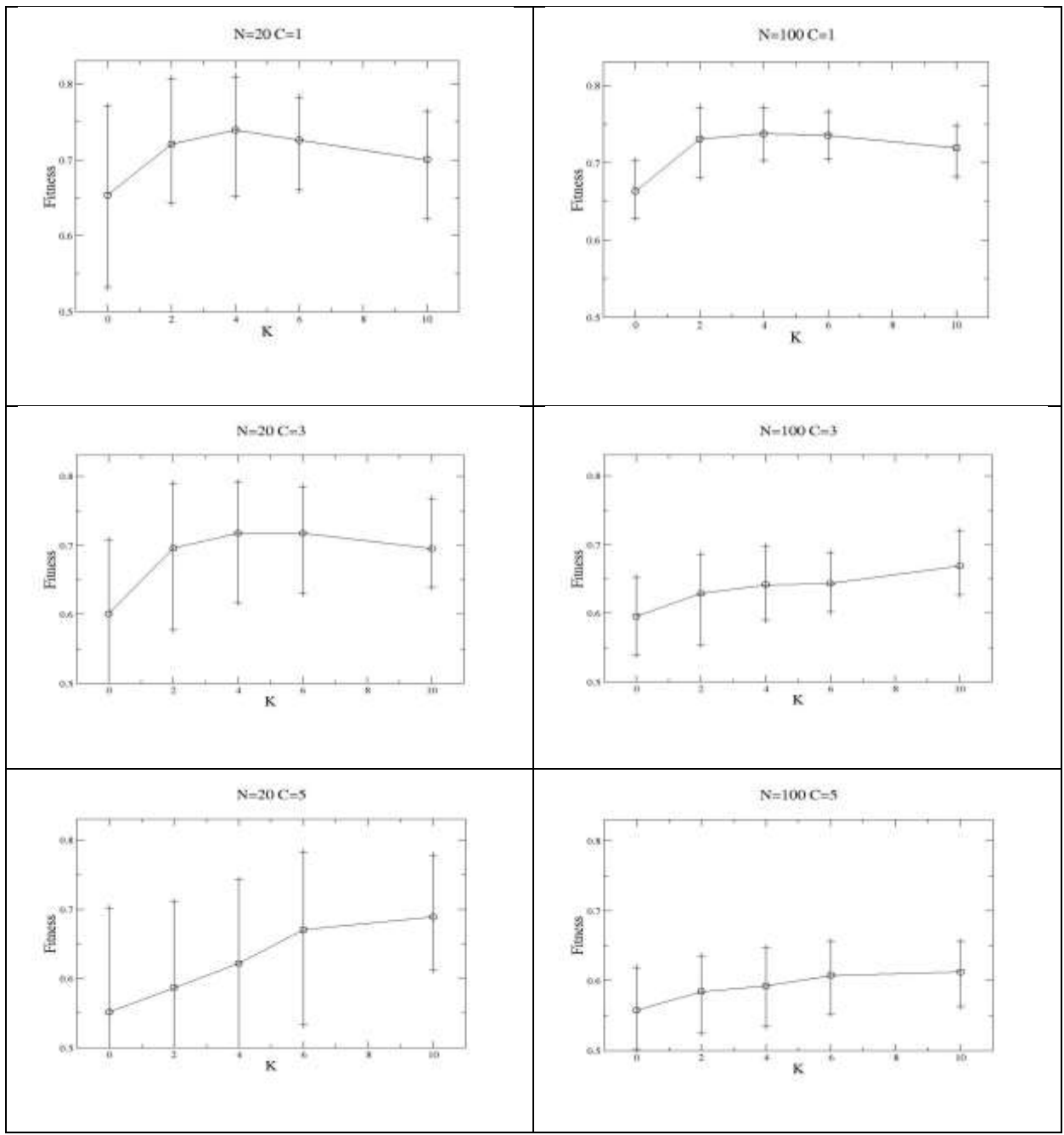

Figure 7: Showing the fitness reached after 20,000 generations for a sexual diploid species coevolving with an asexual haploid species on landscapes of varying ruggedness ($K$), coupling ($C$), and length ($N$).

Figure 8 shows examples when both species are sexual diploids. Comparing to Figure 7, with $C$=1 or $C$=5 and $N$=20 or $N$=100, there is no difference in fitness compared to the equivalent case coevolving with an asexual haploid for all *K* (T-test, $p \geq 0.05$). When *C*=3 and *N*=20 there is again no difference for all *K* (T-test, $p \geq 0.05$) but with *N*=100 an increase in fitness is seen when *K*>4 (T-test, $p<0.05$).

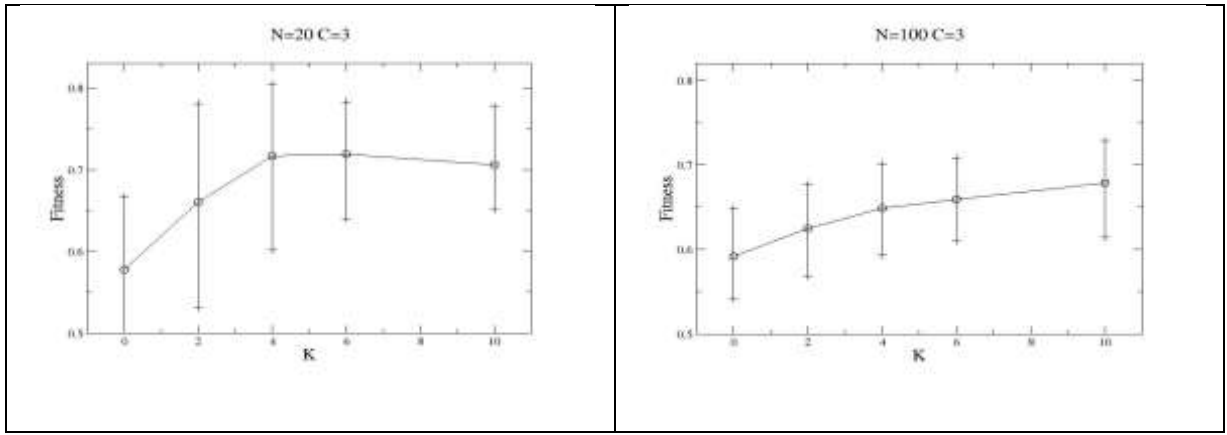

Figure 8: Showing examples of the fitness reached after 20,000 generations for a sexual diploid species coevolving with another sexual diploid species on landscapes of varying ruggedness (*K*) and length (*N*) with *C*=3.

**Reproduction Rates in the NKCS Model**

As in [Kauffman & Johnson, 1992], the above models have assumed all species coevolve at the same rate; each species coevolves in turn. Following [Bull, et al. 2000] who used asexual haploid species, a new parameter *R* can be added to the model to represent the relative rate at which one species evolves - by undertaking *R* rounds of mutation and selection to one round in the other(s). Figure 9 shows how with *N*=100, generally, increasing *R* increases the effects of *C* for a given *K* when *C*>1 with -10<*R*<10. In contrast, no notable effect is seen by increasing *R* for any *K* and *C* values tried when *N*=20 (not shown). It can be noted that *N*=64 (only) was used in [Bull et al., 2000].

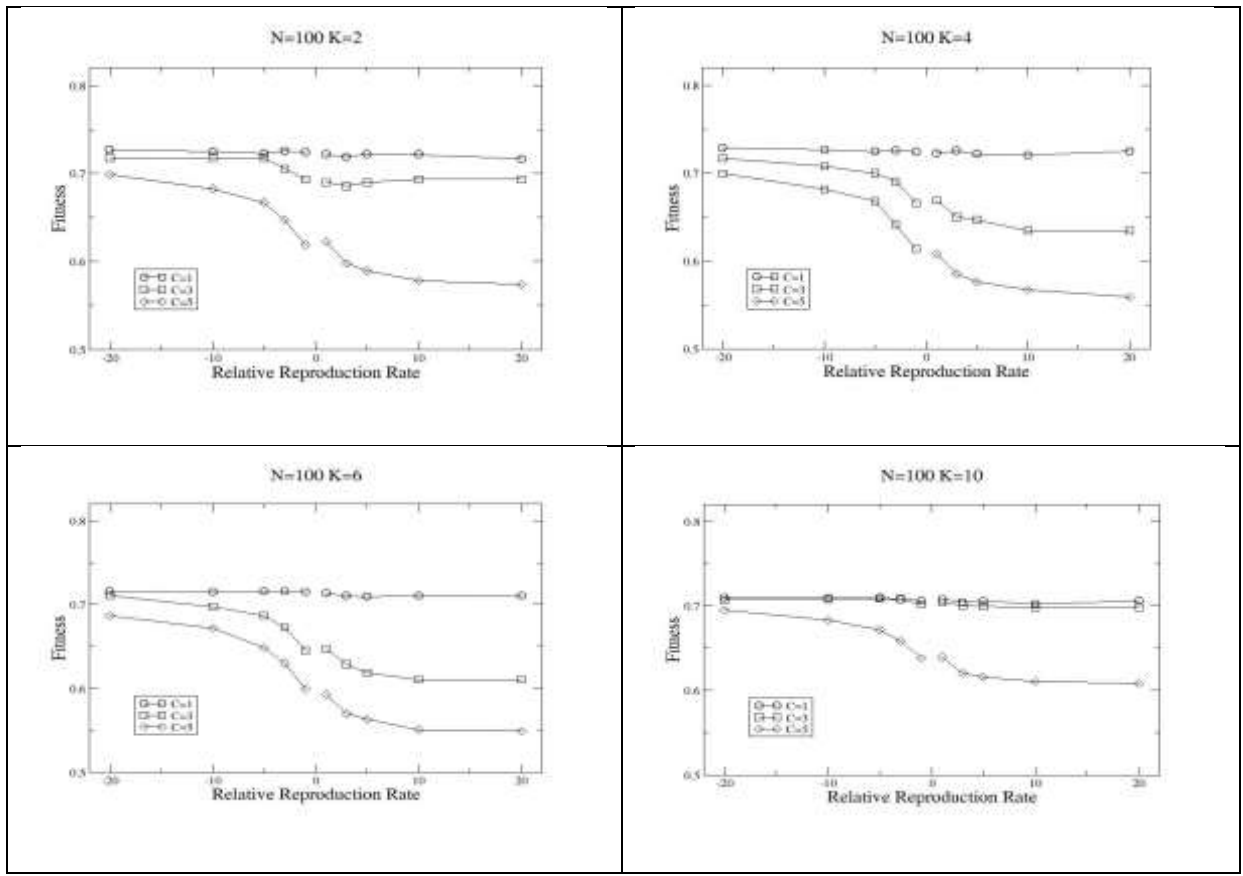

Figure 9: Showing examples of the fitness reached after 20,000 generations for asexual haploid species coevolving with a partner reproducing at a different relative rate $R$. Error bars are not shown for clarity.

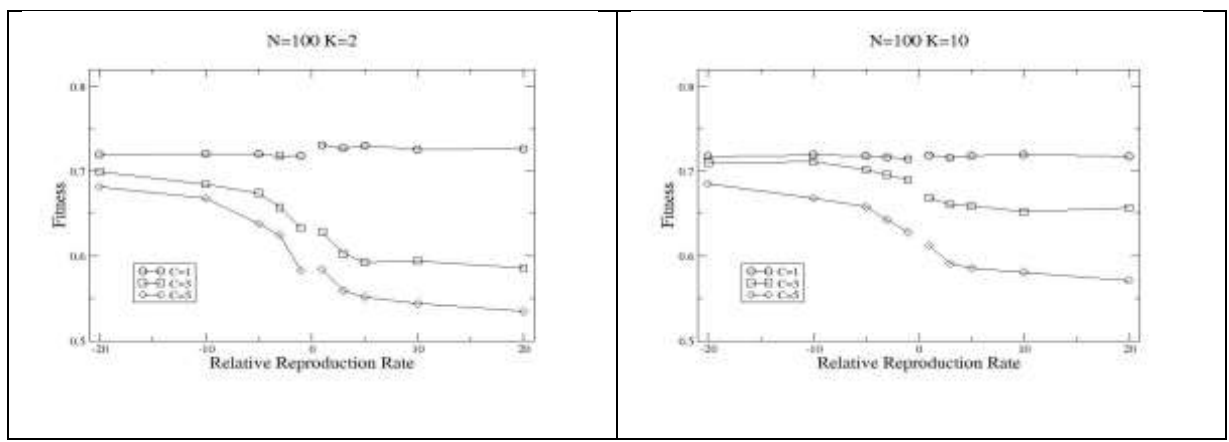

Figure 10: Showing examples of the fitness reached after 20,000 generations for sexual diploid species coevolving with an asexual haploid partner reproducing at a different relative rate $R$.

Figure 10 shows results for a sexual diploid species coevolving with an asexual haploid species. As can be seen, with $N$=100, an increase in $R$ again decreases fitness for all $K$ when $C$>1. Moreover, the fitness reached for $R$>1 is typically significantly lower (T-test, $p$<0.05) than in the equivalent case for the asexual haploid species above (compare to Figure 9) when $C$>1. That is, any advantage (or neutrality) from sexual reproduction seen above with $R$=1 is lost. Results with $N$=20 show no significant change in fitness, as was the case with the asexual haploids (not shown).

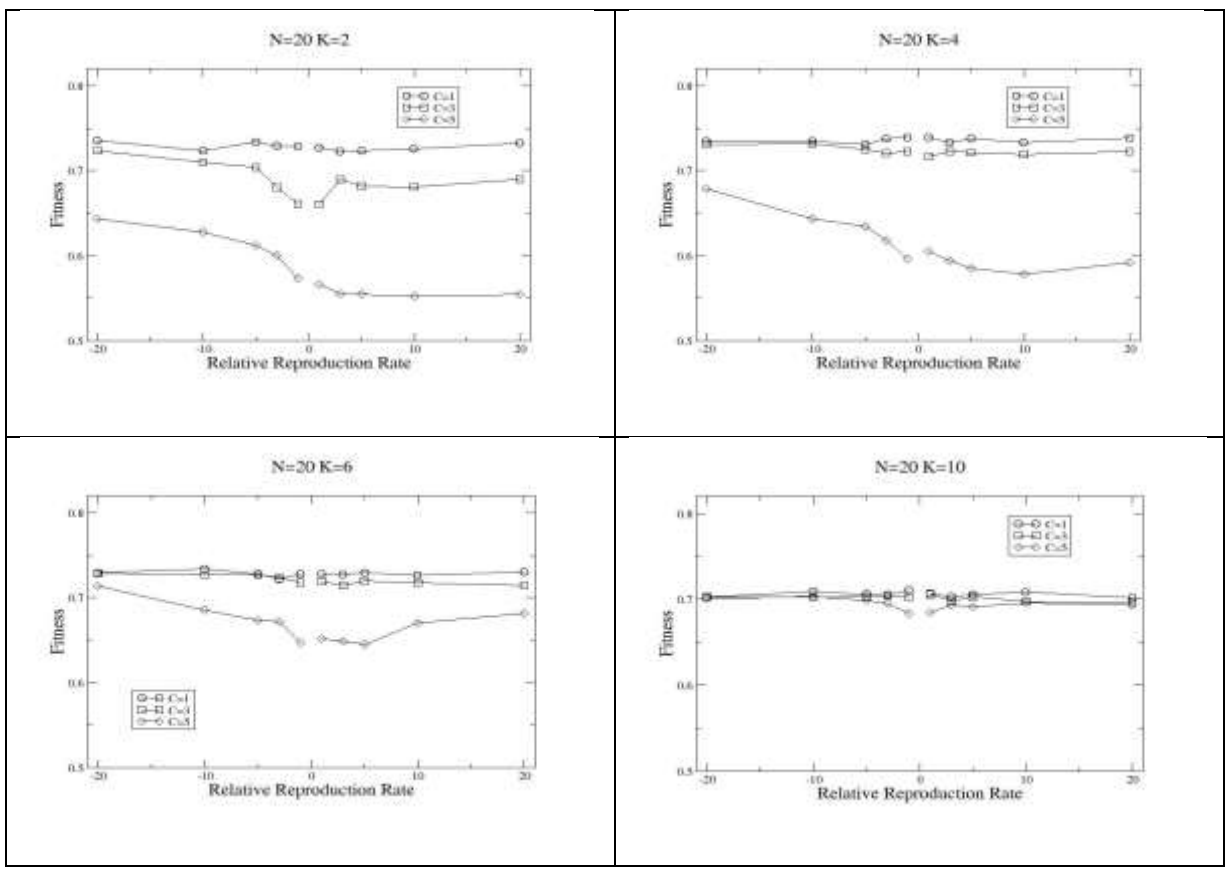

Figure 11: Showing examples of the fitness reached after 20,000 generations for a sexual diploid species coevolving with another sexual diploid species reproducing at a different relative rate $R$.

Finally, Figure 11 shows the case where both species are sexual diploids. As can be seen, with $N$=20, there are instances when increasing $R$ can *increase* the fitness of the slower species, when $K$=2 and $C$=3, when $K$=6 and $C$=5, and when $K$=10 and $C$=5 (T-test, $p$<0.05). Results when $N$=100 are the same as when the partner is asexual above (not shown).

**Conclusion**

Using the NK model, it has recently been suggested that eukaryotic sex exploits a rudimentary form of the Baldwin effect [Bull, 2017]. Following [Bull, 1999], it was shown how the size and ruggedness of the fitness landscape affected the most beneficial amount and frequency of learning. Using the NKCS model, this paper has explored the utility of sex in a coevolutionary context with results showing the learning in two-step meiosis with recombination generally appears more sensitive to coevolution than asexuality. Of particular note is that sex was found to be beneficial for low degrees of fitness landscape coupling and for some cases of high landscape coupling when the partnering species evolves a higher relative rate, where the latter is always highly detrimental for asexual reproduction.